# How do Quantum Effects Influence the Capacitance and Carrier Density of Monolayer MoS₂ Transistors?

Robert K. A. Bennett[1] and Eric Pop[1,*]

[1]*Department of Electrical Engineering, Stanford University, Stanford, California 94305, U.S.A.*

[*]Contact: epop@stanford.edu

**ABSTRACT:** When transistor gate insulators have nanometer-scale equivalent oxide thickness (EOT), the gate capacitance ($C_G$) becomes smaller than the oxide capacitance ($C_{ox}$) due to the quantum capacitance and charge centroid capacitance of the channel. Here, we study the capacitance of monolayer MoS₂ as a prototypical two-dimensional (2D) channel while considering spatial variations in the potential, charge density, and density of states. At 0.5 nm EOT, the monolayer MoS₂ capacitance is smaller than its quantum capacitance, limiting the single-gated $C_G$ of an *n*-type channel to between 63% and 78% of $C_{ox}$ for gate overdrive voltages between 0.5 and 1 V. Despite these limitations, for dual-gated devices, the on-state $C_G$ of monolayer MoS₂ is 50% greater than that of silicon at 0.5 nm EOT and more than three times that of InGaAs at 1 nm EOT, indicating that 2D semiconductors are promising for nanoscale devices at future technology nodes.

Two-dimensional (2D) semiconductors have emerged over the last decade as promising candidates for channel materials in sub-10-nm metal-oxide-semiconductor field-effect transistors (MOSFETs).[1,2] Using 2D monolayer semiconductors in such transistors is appealing from an electrostatic perspective because their ultrathin channel (< 1 nm) reduces the impact of lateral fringing fields while increasing the 2D semiconductor's out-of-plane capacitance $C_{sc}$ (sometimes called the inversion layer capacitance in bulk semiconductor transistors in the on-state). Although conventional bulk semiconductors (like silicon) suffer from mobility degradation as their channel thickness is reduced to a few nanometers, 2D semiconductors maintain good carrier mobilities even at their monolayer limit, allowing them to simultaneously offer excellent electrostatic control and good on-state conductance.[3,4]

In a field-effect transistor, the total gate capacitance $C_G = q\partial n_{ch}/\partial V_G$ [where $q$ is the elementary charge, $n_{ch}$ is the number of charge carriers (electrons or holes) per unit area, and $V_G$ is the gate voltage] is given by the series capacitance of $C_{sc}$ with the gate insulator capacitance, denoted here as $C_{ox}$ (acknowledging that gate insulators may have nitrides or other components),[5,6] as shown in Figure 1a and equation 1:

$$\frac{1}{C_G} = \frac{1}{C_{sc}} + \frac{1}{C_{ox}}. \tag{1}$$

The semiconductor channel's contribution to $C_G$ is negligible when $C_{sc} \gg C_{ox}$, at which point $C_G \approx C_{ox}$ = $\epsilon_{ox}/t_{ox}$, where $\epsilon_{ox}$ and $t_{ox}$ are the insulator's permittivity and thickness, respectively. If $C_{sc}$ is comparable to $C_{ox}$, however, then $C_{sc}$ can limit $C_G$, thereby limiting the maximum carrier densities achievable in the FET on-state.[5] For example, we demonstrate in this work that for monolayer $MoS_2$, $C_{sc}$ is negligible when the gate insulator's equivalent oxide thickness (EOT) is $\geq 2.5$ nm, although the precise EOT at which $C_{sc}$ becomes negligible varies between semiconductors.[5,7-9]

The $C_{sc}$ has two main components: the centroid capacitance (due to the penetration of the charge centroid into the semiconductor channel[5,10]) and the quantum capacitance $C_q$ (due to Fermi level movement with respect to the energy bands in a semiconductor channel with finite density of states[9,11-13]). For 2D semiconductors, the centroid capacitance has often been taken as infinite (implicitly assuming $C_{sc} \approx C_q$), where $C_q$ is evaluated as[12]

$$C_q = 2\pi \frac{q^2 g_s g_v m^*}{h^2} \left(1 + \frac{\exp\left[E_G/(2k_B T)\right]}{2\cosh\left[q\psi_{ch}/(k_B T)\right]}\right)^{-1}, \tag{2}$$

where $g_s = 2$ is the spin degeneracy, $g_v$ is the valley degeneracy (= 2 and 6 for the lower K-valleys and the higher Q-valleys, respectively, of the conduction band in monolayer $MoS_2$), $m^*$ is the effective mass, $h$ is Planck's constant, $E_G$ is the electronic band gap ($\approx 2.2$ eV for monolayer $MoS_2$, depending on its dielectric environment[14]), $k_B$ is the Boltzmann constant, and $T$ is the absolute temperature (here, ~300 K). Above, $\psi_{ch}$ is the channel potential, typically considered without regard to its variation in the channel of 2D semiconductors (i.e., infinite centroid capacitance) but self-consistently treated in this work with spatial variation of charge density. Although $C_q$ is very small in the off-state, in the on-state a large $|\psi_{ch}|$ pushes the Fermi energy into the channel conduction or valence bands, causing the bracketed term in equation (2) to approach unity and saturating $C_q$ to its degenerate value $C_{dq}$, given by[12,15]

$$C_{dq} = 2\pi \frac{q^2 g_s g_v m^*}{h^2}.$$ (3)

Considering only the lowest-energy conduction and valence bands, $C_{dq} \approx 70$ and $200~\mu\text{F/cm}^2$ for $n$- and $p$-type monolayer MoS$_2$, respectively. Although including higher energy bands (e.g., the Q-valley along the T-line[16] in the monolayer MoS$_2$ conduction bands) would enable larger $C_q$, even these lower-bound estimates of $C_{dq}$ greatly exceed $C_{ox}$ for any realistic EOT, leading most studies to neglect $C_{sc}$.

However, previous experimental studies on monolayer semiconductors, including MoS$_2$, MoSe$_2$, WSe$_2$, and black phosphorus, have reported values of $C_q$ and $C_{sc}$ that are much smaller than their respective $C_{dq}$ when the Fermi energy $E_F$ is pushed beyond the band edges.[17-19] Although these smaller-than-anticipated capacitances could be attributed to extrinsic contributions (like defects), recent theoretical work has shown that for monolayer MoS$_2$, other components of $C_{sc}$ could limit it to values much smaller than $C_{dq}$.[20] However, the contribution of non-uniform carrier distributions across a 2D semiconductor's thickness, as well as the impact that this reduced $C_{sc}$ will have on $C_G$, remain largely unexplored.

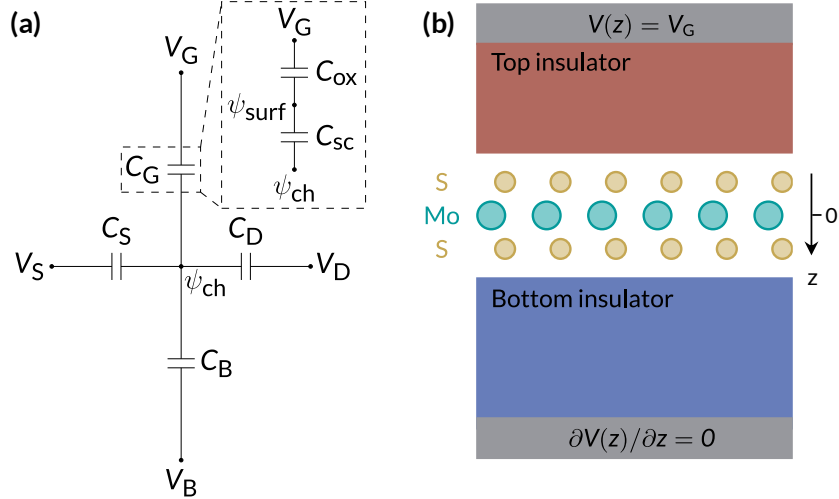

**Figure 1.** (a) Capacitance network model of a 2D transistor. $C$ and $V$ represent capacitances and potentials, subscripts G, S, D, and B denote quantities associated with the gate, source, drain, and the bottom insulator/substrate, respectively. $\psi_{ch}$ is the channel potential, which can vary across the thickness of the channel. The inset shows $C_G$ as the series capacitance of $C_{ox}$ and $C_{sc}$ with an intermediate surface potential $\psi_{surf}$. (b) Schematic of the MoS$_2$-based MOS capacitor considered in this work, along with boundary conditions applied when solving equations (4) and (5).

In this work, we address these gaps in knowledge by self-consistently solving carrier statistics equations with the electrostatic potential distribution across a monolayer MoS$_2$-based MOS capacitor, as shown in Figure 1b. We consider the spatial variation of electrostatic potential $V(z)$ [where $z$ is the cross-plane coordinate labeled in Figure 1b], the volumetric charge density $\rho(z)$, and the local density of states (LDOS)[21,22] across the monolayer thickness, and we write

$$\rho(z) = \mp q \int \text{LDOS}(E, z) \left[ \frac{1}{2} \mp \frac{1}{2} \pm f(E) \right] dE$$ (4)

where $E$ is the energy, $f(E)$ is the Fermi-Dirac distribution, upper (lower) signs are for electrons (holes), and the channel carrier density $n_{ch}$ is obtained by integrating $|\rho(z)|/q$. Applying a gate voltage changes $\rho(z)$ and $n_{ch}$ by modulating the local electrostatic potential $V(z)$, pushing $E_F$ from mid-gap towards the conduction (valence) bands and populating the channel with electrons (holes).

Here, we assume that the intrinsic semiconductor $E_F$ is at the mid-gap energy, allowing us to equate $E_F$ with $qV(z)$ when computing $f(E)$ in equation (4), where $qV(z)$ is also referenced to mid-gap. As both $\rho(z)$ and $V(z)$ are unknowns, we solve equation (4) self-consistently with Poisson's equation,

$$\frac{\partial}{\partial z}\left[\epsilon(z)\frac{\partial V}{\partial z}\right] = -\rho(z),$$ (5)

where $\epsilon(z)$ is the permittivity. We discretize $V(z)$ and $\rho(z)$ along a one-dimensional grid, including the gate voltage boundary condition $[V(z) = V_G]$ at the top of the gate insulator and a Neumann boundary condition $[\partial V(z)/\partial z = 0]$ at the opposite side of the bottom insulator in the structure shown in Figure 1b.

We model $\epsilon(z)$ as a step-like function that transitions from $\epsilon_{ox}$ to the MoS$_2$ permittivity[17] $4\epsilon_0$ (where $\epsilon_0$ is the permittivity of vacuum) at $z = -t_{ch}/2$, then to the permittivity of SiO$_2$ ($3.9\epsilon_0$) at $z = t_{ch}/2$, where $t_{ch} = 0.615$ nm is the MoS$_2$ monolayer thickness. We note that this dielectric profile is approximate; many different estimates for the out-of-plane dielectric constant of monolayer MoS$_2$ have been reported,[4,17,23,24] and it is unclear how $\epsilon(z)$ varies spatially at the insulator/MoS$_2$ interface. Furthermore, it is uncertain if $\epsilon(z)$ is mostly constant across the thickness of the MoS$_2$ monolayer or if, like graphene,[25] $\epsilon(z)$ is a function of position within the monolayer. Once these factors are known, they can be incorporated into our model by substituting the appropriate dielectric profile into equation (5).

We extract the LDOS of monolayer MoS$_2$ using density functional theory (DFT) with Quantum ESPRESSO software.[26] All calculations are performed on a $151 \times 151 \times 1$ $k$-point grid using projector-augmented wave pseudopotentials with kinetic energy cutoffs of 60 Ry for wave functions and 480 Ry for charge densities and potentials. After computing the LDOS for a primitive cell in three dimensions, we average the LDOS across the in-plane directions to represent it only as functions of $E$ and $z$. Then, to ensure that the LDOS at a specific energy will always sum to the magnitude of the DOS at that same energy, we express it as

$$\text{LDOS}(E, z) = L(E, z)\text{DOS}(E),$$ (6)

where $L(E,z)$ is the spatial distribution of the LDOS,[21] which we normalize at each energy level $E_n$:

$$L(E = E_n, z) = \frac{\text{LDOS}(E = E_n, z)}{\int_{-\infty}^{\infty}\text{LDOS}(E = E_n, z)\mathrm{d}z}.$$ (7)

Figures 2a,b show $L(E,z)$ in the conduction and valence bands, respectively. At $E \leq E_C + 0.25$ eV (where $E_C$ is the conduction band minimum), the LDOS is confined close to the center of the MoS$_2$ monolayer (i.e., near the Mo atoms) with sharp, narrow peaks appearing just to the left and right of the main central peak. These sharp satellite peaks arise from the spatial distribution of the $4d_{z^2}$ orbital of Mo, which has been shown to dominate the DOS at these energies in previous studies.[27,28] Furthermore, we find that broad peaks centered close to the S atoms appear in the LDOS at $E > E_C + 0.25$ eV. As shown in the projected DOS (pDOS) in Figure 2c, the S atoms begin to contribute to the DOS in the conduction bands at ~0.26 eV above the conduction band minimum, corresponding to the Q-K valley separation $\Delta E_{QK}$ from our DFT simulations. Projected band structures have also shown that S atoms contribute weakly to the electronic structure of monolayer $MoS_2$ at the conduction band minimum, but they contribute noticeably to the Q-valley.[27,29] We therefore attribute these peaks to contributions to the DOS from the S atoms.

Similar laterally positioned peaks are present at all values of $E \leq E_V$ (where $E_V$ is the valence band maximum) we consider in Figure 2b, which is also consistent with the pDOS in the valence bands: as shown in Figure 2d, S atoms contribute to the DOS in the valence bands at all considered energies. Similarly, projected band structures have shown that both Mo and S atoms contribute significantly to the valence bands of monolayer $MoS_2$.[27,29]

We note that the exact $\Delta E_{QK}$ for monolayer $MoS_2$ in vacuum is not precisely known[30] and that the band structure of monolayer $MoS_2$ can vary depending on strain[31] and its surrounding dielectric environment.[14] For example, the experimental $\Delta E_{QK}$ for monolayer $MoS_2$ encased in quartz and $WS_2$ is $\Delta E_{QK} \approx 0.11$ eV,[32] and simulated values range between 0.071 and 0.270 eV, depending on the approach used.[30] To accommodate this uncertainty in the value of $\Delta E_{QK}$, we investigate its effect on $C_{sc}$ and $n_{ch}$ in Section S1 of the Supporting Information. We also note from Figures 2a,b that the LDOS extends slightly beyond $t_{ch} = 0.615$ nm, which occurs because DFT simulations of 2D $MoS_2$ assume that the semiconductor is surrounded by vacuum; in reality, a semiconductor's LDOS cannot so easily penetrate into an insulator.[33] However, we shortly demonstrate that this non-ideality should not significantly affect our results.

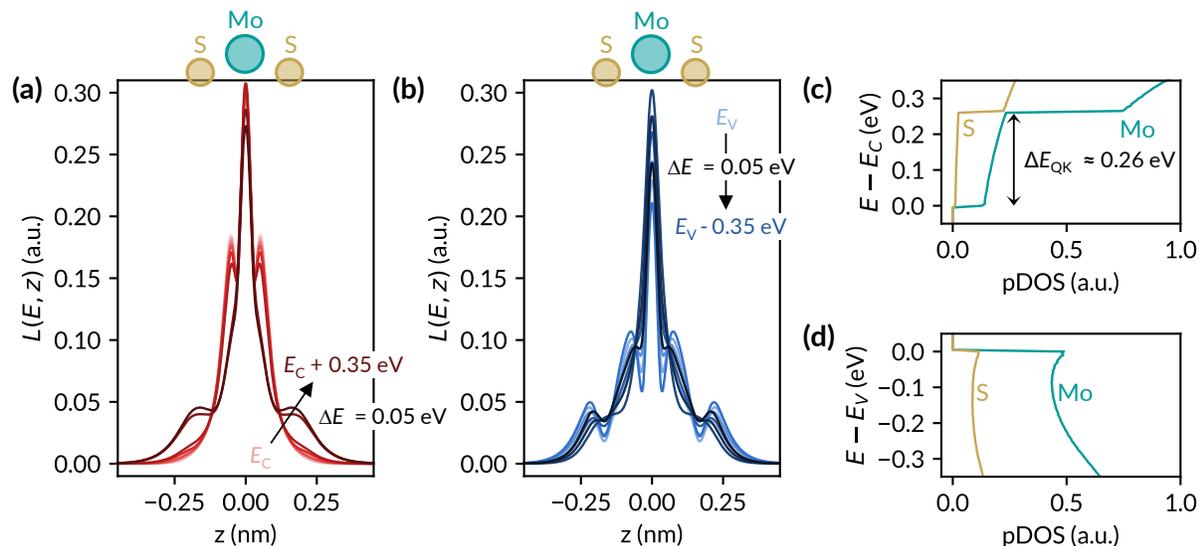

**Figure 2.** Normalized spatial distribution of the local density of states $L(E,z)$ across monolayer $MoS_2$ at (a) $E \geq E_C$, and (b) $E \leq E_V$, where the $z$ coordinates of Mo and S atoms align with the location of atoms at the top of the figures (size of atoms are not to scale). Projected density of states (pDOS) for (c) conduction bands and (d) valence bands of monolayer $MoS_2$, where the contributions of all orbitals from each individual atom are summed together. Note that the contributions from only one S atom are shown in (c) and (d); a.u., arbitrary units.

In Figures 3a,b, we calculate and plot $C_{sc}$ of monolayer $MoS_2$ on linear and logarithmic y-axes, respectively, by self-consistently solving equations (4) and (5) under an applied gate bias, with the corresponding $n_{ch}$ values plotted in Figure 3c. For now, we set the permittivity of the gate insulator to an extremely large value ($C_{ox} \rightarrow \infty$), so that $\psi_{surf} = V_G$, allowing us to study the intrinsic capacitance of monolayer $MoS_2$ by neglecting the potential drop across the gate insulator. We will shortly relax this assumption and study monolayer $MoS_2$-based capacitors with finite EOTs.

At $|\psi_{surf}| - E_G/(2q) < 0.25$ V, the capacitance of $p$-type $MoS_2$ exceeds that of $n$-type $MoS_2$, which is due to the DOS near the valence band edge being larger than the DOS near the conduction band edge (Figures 2c,d). However, as $\psi_{surf}$ is pushed farther into the conduction bands, the slopes of both $C_{sc}$ and $n_{ch}$ increase sharply for $n$-type $MoS_2$. This increase is due to the step-like increase in the DOS at the Q-valley, noting that thermal broadening in equation (4) allows the DOS from the Q-valley to also contribute when $E_F$ is a few $k_BT$ below the edge of the Q-valley. We note that this effect has been experimentally observed in $MoSe_2$ and $WSe_2$ monolayers, which have similar electronic structures to that of $MoS_2$ (including a Q-valley above the conduction band edge). Thus, the shape of the electron density in our Figure 3c resembles similar experimental curves for $MoSe_2$ and $WSe_2$ monolayers.[18]

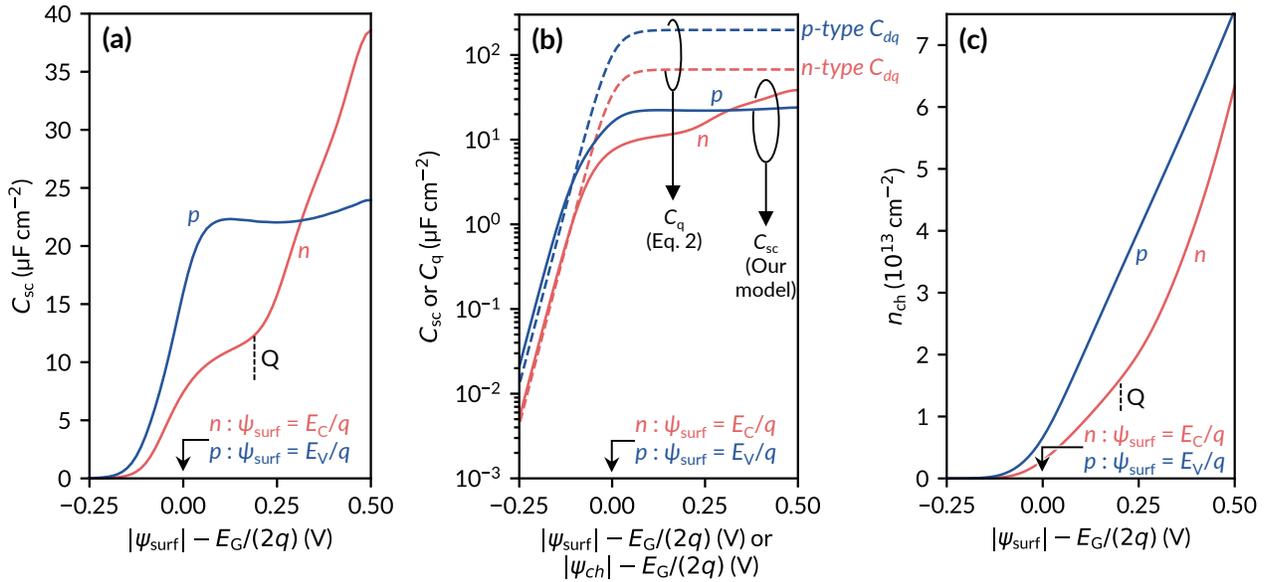

**Figure 3.** Semiconductor capacitance $C_{sc}$ as a function of $|\psi_{surf}| - E_G/(2q)$ plotted on (a) linear and (b) logarithmic axes. For comparison, we also plot quantum capacitance $C_q$ as a function of the channel potential $\psi_{ch}$ [calculated using equation (2) with $g_v m^* = 1.01 m_0$ and $2.96 m_0$ for electrons and holes,[4] respectively] in (b). (c) Charge carrier density $n_{ch}$ for $n$-type and $p$-type monolayer $MoS_2$. The approximate location at which the Q-valley begins to contribute to the capacitance and carrier density of $n$-type monolayer $MoS_2$ is marked in (a) and (c) with a vertical dashed line. Note, carrier densities over ~$2 \times 10^{13}$ cm$^{-2}$ are difficult to access in practice with conventional dielectrics (e.g. $HfO_2$) but can be accessed with solid or liquid electrolyte gating.

We also compare our computed $C_{sc}$ values to the conventional $C_q$ for both $n$- and $p$-type monolayer $MoS_2$, which we have plotted alongside $C_{sc}$ in Figure 3b. Although our calculated $C_{sc}$ closely matches $C_q$ at small gate voltage (i.e., non-degenerate surface potentials), we find that at high gate voltages (i.e., degenerate potentials), our computed $C_{sc}$ values are substantially lower than the traditionally calculated $C_q = C_{dq} = 70$ μF/cm$^2$ for $n$-type and $200$ μF/cm$^2$ for $p$-type monolayer $MoS_2$, respectively. To understand why $C_{sc}$ matches $C_q$ only for *non-degenerate* potentials, we first plot the charge distributions and potential across the thickness of $n$-type ($p$-type) $MoS_2$ when $\psi_{surf}$ is 0.3 V below (above) the conduction (valence) band edge in Figures 4a,b. For this non-degenerate case, the charge distributions are nearly symmetric across the channel, closely matching the LDOS distributions shown in Figures 2a,b. This result is consistent with the potential profile shown in Figure 4b: there is nearly no potential drop across the monolayer $MoS_2$ thickness, allowing electronic states to contribute to the carrier density equally efficiently, regardless of their location in the channel. Hence, $C_{sc} \approx C_q$ in this regime.

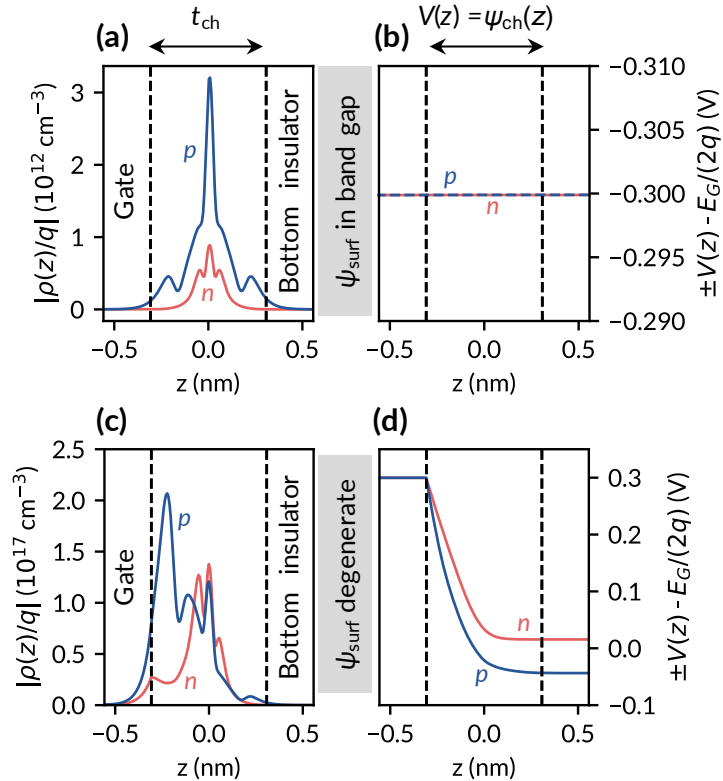

**Figure 4.** (a) Distributions of charge density and (b) potential profile across the $MoS_2$ monolayer with $\psi_{surf}$ inside the band gap, 0.3 eV below the conduction band edge (for electrons) and 0.3 eV above the valence band edge (for holes). (c) Distributions of charge density and (d) potential profile across the $MoS_2$ monolayer with $\psi_{surf}$ of 0.3 eV above the conduction band edge (for electrons) and 0.3 eV below the valence band edge (for holes). Dashed lines indicate boundaries with the gate electrode and bottom insulator. Note, the carrier densities in (c) are much greater than in (a). In (c), the carrier distribution also extends slightly outside $t_{ch}$ by < 10 % of total $n_{ch}$, meaning that the charge centroid is closer to the gate[20] and that $C_{sc}$ is slightly overestimated in this study. Finally, we note that the potential drops are different for $n$- and $p$-type devices in Figure 4d because the capacitance of $n$-type monolayer $MoS_2$ is smaller than that of $p$-type $MoS_2$ for the $\psi_{surf}$ we consider here.[34]

Next, to understand why $C_{sc} < C_q$ at *degenerate* potentials, we plot the charge distributions and potential across the thickness of monolayer $n$-type ($p$-type) MoS$_2$ when $\psi_{surf}$ is 0.3 V above (below) the conduction (valence) band edge in Figures 4c,d. We find that for this degenerate case, the charge distributions are heavily asymmetric and skewed towards the gate electrode. This asymmetry can be explained from the potential drops in Figure 4d, which shows that the local electrostatic potential is highest near the gate electrode and rapidly decays across the monolayer channel. We recall from Figures 2a,b that most available states are near the center of the channel in this region of relatively low potential. Therefore, many states are unable to efficiently contribute to $n_{ch}$, which is why $C_{sc} < C_{dq}$ even at high $\psi_{surf}$. As a result, the charge density is askew across the thickness of the MoS$_2$ monolayer, with the S atoms opposite to the gate contributing little to the channel charge. We conclude that at degenerate surface potentials, the shapes of the LDOS and potential profile play pivotal roles in dictating $C_{sc}$ for 2D semiconductors.

We next consider the impact of $C_{sc}$ in practical MOS devices based on monolayer MoS$_2$. Figures 5a,b plot $C_G$ and $n_{ch}$ for MOS capacitors as in Figure 1b with EOTs of 0.5, 1, and 2.5 nm ($\epsilon_{ox} = 20\epsilon_0$). For comparison, we additionally plot $C_{ox} = 3.9\epsilon_0/\text{EOT}$, as well as the classical carrier density $n_{ch}^{classical} = (V_G - V_T)C_{ox}/q$ (for electrons; the bracketed term is negated for holes), where the threshold voltage is $V_T = \pm E_G/(2q)$ for $n$- and $p$-type devices, assuming the same gate metal. At EOT = 2.5 nm, $C_G$ saturates close to $C_{ox}$ and $n_{ch} \approx n_{ch}^{classical}$, but the observed $C_G$ and $n_{ch}$ deviate significantly from $C_{ox}$ and $n_{ch}^{classical}$, respectively, at EOTs of 0.5 and 1 nm. For example, at EOT = 0.5 nm, the $C_G$ of $n$-type MoS$_2$ increases from 4.38 μF/cm$^2$ at 0.5 V to 5.35 μF/cm$^2$ at 1 V, remaining between 63% and 78% of $C_{ox}$. However, we note that $C_G$ is sensitive to small variations in $V_G$ and to the position of the Q-valley for $n$-type MoS$_2$, as captured in Figure 5a and discussed in Supporting Information Section S1.

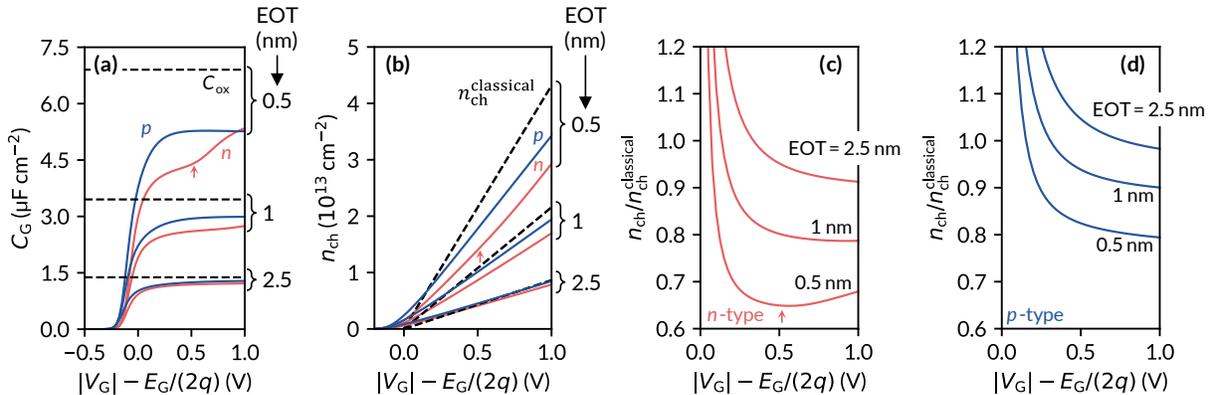

**Figure 5.** (a) Gate capacitance $C_G$ and (b) charge carrier density $n_{ch}$ for $n$-type and $p$-type monolayer MoS$_2$ as functions of $V_G$ at EOT = 0.5, 1, and 2.5 nm. Solid red (blue) lines represent $n$-type ($p$-type) MoS$_2$. Dashed lines mark the oxide capacitance $C_{ox}$ and the conventionally calculated carrier density $n_{ch}^{classical}$, highlighting the importance of quantum corrections in the limit of ultrathin EOT. Small red arrows mark approximately where the Q-valley of $n$-type MoS$_2$ begins to contribute. (c) Calculated $n_{ch}/n_{ch}^{classical}$ for $n$-type MoS$_2$ and (d) $p$-type MoS$_2$. The true charge density is lower than the classical estimate in the limit of high $V_G$ when transistors are strongly turned on. However, $n_{ch}^{classical}$ approaches zero and underestimates the true charge density near threshold.

Similarly, we find that the classical equation overestimates $n_{ch}$ in 2D MOS capacitors with small EOTs, as shown in Figures 5c,d for $n$- and $p$-type MoS$_2$, respectively. At EOTs of 0.5 and 1 nm, the channel carrier density $n_{ch}$ is as small as 65% and 79% (79% and 89%) of $n_{ch}^{classical}$ for $n$-type ($p$-type) MoS$_2$ in the $V_G$ range considered here. We note that contributions from the Q-valley are visible for the $n$-type device with EOT = 0.5 nm, leading to an increase in $C_G$ and $n_{ch}$ when the voltage is sufficiently high. This effect has also been observed experimentally in ion-gated MoSe$_2$ and WSe$_2$ monolayers, which have similar band structures to that of monolayer MoS$_2$.[18]

We note that $n_{ch}^{classical} = C_{ox}(V_G - V_T)/q$ should not be applied near or below $V_T$, because this expression neglects subthreshold charge.[6] Instead, $n_{ch}$ may be approximated in both the off- and on-states by taking $C_G$ as the series combination of $C_q$ and $C_{ox}$,[11,12] and then integrating the result (up to the relevant $V_G$) to find the carrier density. However, as we demonstrate in Section S2 of the Supporting Information, correcting for $C_q$ in this manner still significantly overestimates both $C_G$ and $n_{ch}$ in the on-state for 2D channels with low EOT, highlighting the importance of including both quantum and centroid effects when modeling these devices.

Finally, we assess how $C_{sc}$ limits $C_G$ for monolayer MoS$_2$ compared to other semiconductors, including silicon. Although decreasing $t_{ch}$ improves the channel electrostatics, $t_{ch} < 5$ nm causes surface scattering to limit silicon carrier mobilities.[2,35,36] At the limit $t_{ch} \approx 5$ nm, a previous study[37] has shown that the $C_{sc}$ of silicon limits a dual-gated silicon FET with an EOT of 0.5 nm to $C_G \approx 7$ µF/cm$^2$ at an overdrive $V_{OV} = V_G - V_T = 1$ V. In direct comparison, our calculations show that a similar dual-gated structure with $n$-type monolayer MoS$_2$ offers $C_G \approx 10.9$ µF/cm$^2$ (10.6 µF/cm$^2$ for $p$-type), over 50% greater than that of silicon. The MoS$_2$ advantage persists even when the silicon thickness is reduced[37] to 2.5 nm, which yields $C_G \approx 8.1$ µF/cm$^2$. We refer the reader to Section S3 of the Supporting Information for a description of how we calculated $C_G$ and $n_{ch}$ for dual-gated devices and for full $C_G$ and $n_{ch}$ vs. $V_G$ curves.

The $C_{sc}$ of monolayer MoS$_2$ compares even more favorably to III-V semiconductors, whose low DOS are known to limit their $C_q$. A previous study[38] has shown that $C_q$ limits dual-gated InGaAs MOS capacitors with EOT = 1 nm to $C_G < 1.6$ µF/cm$^2$ at channel thickness $t_{ch} = 25$ nm; as $t_{ch}$ decreases, this $C_G$ worsens because the DOS shrinks due to quantum confinement effects.[38] Using the same approach as above, we find that a dual-gated monolayer $n$-type MoS$_2$ capacitor with EOT = 1 nm offers $C_G \approx 5.55$ µF/cm$^2$ (or 6.00 µF/cm$^2$ for $p$-type) at $V_{OV} = 1$ V, over three times higher than InGaAs. The results from Figure 5a also indicate that single-gated monolayer MoS$_2$ capacitors with EOT of 2.5 nm offer higher $C_G$ than those reported for single-gated In$_{0.7}$Ga$_{0.3}$As and InAs capacitors with similar or lower EOTs.[5]

In conclusion, we have shown that variations in carrier density (i.e., the centroid capacitance), potential, and density of states across the thickness of monolayer MoS$_2$ severely limits its on-state $C_{sc}$ to values well below its degenerate quantum capacitance. As a result, gate capacitance estimates made by classical equations must be corrected when evaluating the carrier density and mobility of devices with small EOTs below ~2 nm. Nevertheless, we find that in strong inversion, the $C_G$ of dual-gated $n$-type monolayer

MoS$_2$ capacitors is over 50% higher than dual-gated silicon MOS capacitors at EOT = 0.5 nm and over three times higher than InGaAs capacitors at EOT = 1 nm. The monolayer MoS$_2$ capacitance advantage is higher at lower EOTs, ultimately indicating that strong current and transconductance may be achieved in such 2D transistors if their channel mobility and contact resistance continue to be improved.

## ASSOCIATED CONTENT

**Supporting Information**: Description and variation of the Q-K energy valley separation; comparison to quantum capacitance approximation; methodology and results for dual-gated calculations.

**Author contributions**: R.K.A.B. and E.P. conceived the idea and wrote the manuscript. R.K.A.B. carried out all calculations.

**Notes**: The authors declare no competing financial interests.

## ACKNOWLEDGMENTS

R.K.A.B. acknowledges support from the Stanford Graduate Fellowship (SGF) and the NSERC PGS-D programs. The authors also acknowledge partial support from the Stanford SystemX Alliance and from ASCENT, one of six centers in JUMP, an SRC program sponsored by DARPA.

# REFERENCES

1    Das, S. *et al.* Transistors based on two-dimensional materials for future integrated circuits. *Nat Electronics* **4**, 786-799 (2021). DOI: 10.1038/s41928-021-00670-1

2    Liu, Y. *et al.* Promises and prospects of two-dimensional transistors. *Nat* **591**, 43-53 (2021). DOI: 10.1038/s41586-021-03339-z

3    Song, C. *et al.* Growth and Interlayer Engineering of 2D Layered Semiconductors for Future Electronics. *ACS Nano* **14**, 16266-16300 (2020). DOI: 10.1021/acsnano.0c06607

4    Klinkert, C. *et al.* 2-D Materials for Ultrascaled Field-Effect Transistors: One Hundred Candidates under the Ab Initio Microscope. *ACS Nano* **14**, 8605-8615 (2020). DOI: 10.1021/acsnano.0c02983

5    Donghyun, J., Kim, D., Taewoo, K. & Alamo, J. A. d. Quantum capacitance in scaled down III–V FETs. *2009 IEEE Int Electron Devices Meeting (IEDM)*, 1-4 (2009). DOI: 10.1109/IEDM.2009.5424312

6    Lundstrom, M. S. Fundamentals of Nanotransistors (World Scientific Publishing Company, 2017, Singapore), p. 97, 128.

7    Ganeriwala, M. D. *et al.* Modeling of Charge and Quantum Capacitance in Low Effective Mass III-V FinFETs. *IEEE J Electron Dev Soc* **4**, 396-401 (2016). DOI: 10.1109/JEDS.2016.2586116

8    Yadav, C., Ganeriwala, M. D., Mohapatra, N. R., Agarwal, A. & Chauhan, Y. S. Compact Modeling of Gate Capacitance in III–V Channel Quadruple-Gate FETs. *IEEE Trans Nanotechnol* **16**, 703-710 (2017). DOI: 10.1109/TNANO.2017.2709752

9    Ilatikhameneh, H. *et al.* Saving Moore's Law Down To 1 nm Channels With Anisotropic Effective Mass. *Sci Rep* **6**, 31501 (2016). DOI: 10.1038/srep31501

10   Pal, H. S., Cantley, K. D., Ahmed, S. S. & Lundstrom, M. S. Influence of Bandstructure and Channel Structure on the Inversion Layer Capacitance of Silicon and GaAs MOSFETs. *IEEE Trans Electron Devices* **55**, 904-908 (2008). DOI: 10.1109/TED.2007.914830

11   Yoon, Y., Ganapathi, K. & Salahuddin, S. How Good Can Monolayer $MoS_2$ Transistors Be? *Nano Lett* **11**, 3768-3773 (2011). DOI: 10.1021/nl2018178

12   Ma, N. & Jena, D. Carrier statistics and quantum capacitance effects on mobility extraction in two-dimensional crystal semiconductor field-effect transistors. *2D Mater* **2**, 015003 (2015). DOI: 10.1088/2053-1583/2/1/015003

13   Wu, P. & Appenzeller, J. Artificial Sub-60 Millivolts/Decade Switching in a Metal–Insulator–Metal–Insulator–Semiconductor Transistor without a Ferroelectric Component. *ACS Nano* **15**, 5158-5164 (2021). DOI: 10.1021/acsnano.0c10344

14   Ryou, J., Kim, Y.-S., Kc, S. & Cho, K. Monolayer $MoS_2$ Bandgap Modulation by Dielectric Environments and Tunable Bandgap Transistors. *Sci Rep* **6**, 29184 (2016). DOI: 10.1038/srep29184

15   Bera, M. K. *et al.* Influence of Quantum Capacitance on Charge Carrier Density Estimation in a Nanoscale Field-Effect Transistor with a Channel Based on a Monolayer $WSe_2$ Two-Dimensional Crystal Semiconductor. *J Electron Mater* **48**, 3504-3513 (2019). DOI: 10.1007/s11664-019-07058-0

16   Rasmussen, F. A. & Thygesen, K. S. Computational 2D Materials Database: Electronic Structure of Transition-Metal Dichalcogenides and Oxides. *J Phys Chem C* **119**, 13169-13183 (2015). DOI: 10.1021/acs.jpcc.5b02950

17   Chen, X. *et al.* Probing the electron states and metal-insulator transition mechanisms in molybdenum disulphide vertical heterostructures. *Nat Commun* **6**, 6088 (2015). DOI: 10.1038/ncomms7088

18   Zhang, H., Berthod, C., Berger, H., Giamarchi, T. & Morpurgo, A. F. Band Filling and Cross Quantum Capacitance in Ion-Gated Semiconducting Transition Metal Dichalcogenide Monolayers. *Nano Lett* **19**, 8836-8845 (2019). DOI: 10.1021/acs.nanolett.9b03667

19   Chen, X. *et al.* Probing the electronic states and impurity effects in black phosphorus vertical heterostructures. *2D Mater* **3**, 015012 (2016). DOI: 10.1088/2053-1583/3/1/015012

20   Fang, N. & Nagashio, K. Quantum-mechanical effect in atomically thin $MoS_2$ FET. *2D Mater* **7**, 014001 (2019). DOI: 10.1088/2053-1583/ab42c0

21   Li, W. *et al.* Density of States and Its Local Fluctuations Determined by Capacitance of Strongly Disordered Graphene. *Sci Rep* **3**, 1772 (2013). DOI: 10.1038/srep01772

22    Rickhaus, P. *et al.* The electronic thickness of graphene. *Sci Adv* **6**, eaay8409 DOI: 10.1126/sciadv.aay8409

23    Laturia, A., Van de Put, M. L. & Vandenberghe, W. G. Dielectric properties of hexagonal boron nitride and transition metal dichalcogenides: from monolayer to bulk. *npj 2D Mater and Appl* **2**, 6 (2018). DOI: 10.1038/s41699-018-0050-x

24    Lu, A. K. A., Pourtois, G., Luisier, M., Radu, I. P. & Houssa, M. On the electrostatic control achieved in transistors based on multilayered $MoS_2$: A first-principles study. *J Appl Phys* **121**, 044505 (2017). DOI: 10.1063/1.4974960

25    Fang, J., Vandenberghe, W. G. & Fischetti, M. V. Microscopic dielectric permittivities of graphene nanoribbons and graphene. *Phys Rev B* **94**, 045318 (2016). DOI: 10.1103/PhysRevB.94.045318

26    Giannozzi, P. *et al.* QUANTUM ESPRESSO: a modular and open-source software project for quantum simulations of materials. *J Phys Condens Mat* **21**, 395502 (2009). DOI: 10.1088/0953-8984/21/39/395502

27    Padilha, J. E., Peelaers, H., Janotti, A. & Van de Walle, C. G. Nature and evolution of the band-edge states in $MoS_2$: From monolayer to bulk. *Phys Rev B* **90**, 205420 (2014). DOI: 10.1103/PhysRevB.90.205420

28    Sun, X., Wang, Z., Li, Z. & Fu, Y. Q. Origin of Structural Transformation in Mono- and Bi-Layered Molybdenum Disulfide. *Sci Rep* **6**, 26666 (2016). DOI: 10.1038/srep26666

29    Kormányos, A. *et al.* k·p theory for two-dimensional transition metal dichalcogenide semiconductors. *2D Mater* **2**, 022001 (2015). DOI: 10.1088/2053-1583/2/2/022001

30    Gaddemane, G., Gopalan, S., Van de Put, M. L. & Fischetti, M. V. Limitations of ab initio methods to predict the electronic-transport properties of two-dimensional semiconductors: the computational example of 2H-phase transition metal dichalcogenides. *J Comput Electron* **20**, 49-59 (2021). DOI: 10.1007/s10825-020-01526-1

31    Datye, I. M. *et al.* Strain-Enhanced Mobility of Monolayer $MoS_2$. *Nano Lett* **22**, 8052–8059 (2022). DOI: 10.1021/acs.nanolett.2c01707

32    Hill, H. M., Rigosi, A. F., Rim, K. T., Flynn, G. W. & Heinz, T. F. Band Alignment in $MoS_2$/$WS_2$ Transition Metal Dichalcogenide Heterostructures Probed by Scanning Tunneling Microscopy and Spectroscopy. *Nano Lett* **16**, 4831-4837 (2016). DOI: 10.1021/acs.nanolett.6b01007

33    Fiore, S., Klinkert, C., Ducry, F., Backman, J. & Luisier, M. Influence of the hBN Dielectric Layers on the Quantum Transport Properties of $MoS_2$ Transistors. *Mater* **15** (2022). DOI: 10.3390/ma15031062

34    In a physical device, the remainder of the potential would be dropped across the bottom insulator, which is not observed here because we apply a Neumann boundary condition at the far side of the bottom insulator. However, we have verified that the potentials across the channel of Figure 4b,d become nearly identical to those obtained when we instead ground the far side of the bottom insulator (note that electrical ground corresponds to the mid-gap voltage) and allow the bottom insulator thickness to approach infinity.

35    Uchida, K. *et al.* Experimental study on carrier transport mechanism in ultrathin-body SOI nand p-MOSFETs with SOI thickness less than 5 nm. *Digest. Int Electron Devices Meeting*, 47-50 (2002). DOI: 10.1109/IEDM.2002.1175776

36    English, C. D., Shine, G., Dorgan, V. E., Saraswat, K. C. & Pop, E. Improved Contacts to $MoS_2$ Transistors by Ultra-High Vacuum Metal Deposition. *Nano Lett* **16**, 3824-3830 (2016). DOI: 10.1021/acs.nanolett.6b01309

37    Khan, A. I., Ashraf, M. K. & Khosru, Q. D. M. Effects of wave function penetration on gate capacitance modeling of nanoscale double gate MOSFETs. *2007 IEEE Conference on Electron Devices and Solid-State Circuits*, 137-140 (2007). DOI: 10.1109/EDSSC.2007.4450081

38    Yadav, C. *et al.* Capacitance Modeling in III–V FinFETs. *IEEE Trans Electron Devices* **62**, 3892-3897 (2015). DOI: 10.1109/TED.2015.2480380

# Supporting Information

# How do Quantum Effects Influence the Capacitance and Carrier Density of Monolayer MoS₂ Transistors?

Robert K. A. Bennett[1] and Eric Pop[1,*]

*Department of Electrical Engineering, Stanford University, Stanford, California 94305, U.S.A.*

*Contact: epop@stanford.edu

## S1. Effect of Q-K Energy Valley Separation

Our density functional theory (DFT) calculations yield an energy difference $\Delta E_{QK} \approx 0.26$ eV between the Q- and K-valleys in the conduction bands of monolayer MoS₂, as labeled in Figure S1a (we note this Q-valley is sometimes called T or Λ). However, the computed value of $\Delta E_{QK}$ for monolayer MoS₂ is highly dependent on the input parameters used in DFT (e.g., exchange-correlation functionals and pseudopotentials),[1] where the settings that yield the actual $\Delta E_{QK}$ of monolayer MoS₂ in vacuum are presently unclear. The band structure of monolayer MoS₂ also varies with the surrounding dielectric environment,[2] further complicating the question of which $\Delta E_{QK}$ is relevant for a given physical system.

To accommodate this uncertainty in the "correct" $\Delta E_{QK}$, we repeat calculations of the semiconductor capacitance $C_{sc}$ and carrier density $n_{ch}$ for $n$-type MoS₂ with $\Delta E_{QK} = 0.13$ eV and compare these results to those obtained using $\Delta E_{QK} = 0.26$ eV in the main text. As $p$-type MoS₂ does not have an associated $\Delta E_{QK}$ or similar low-energy peaks that contribute to its density of states (DOS) in the range of energies of interest, the $p$-type results would be unchanged and are not repeated here.

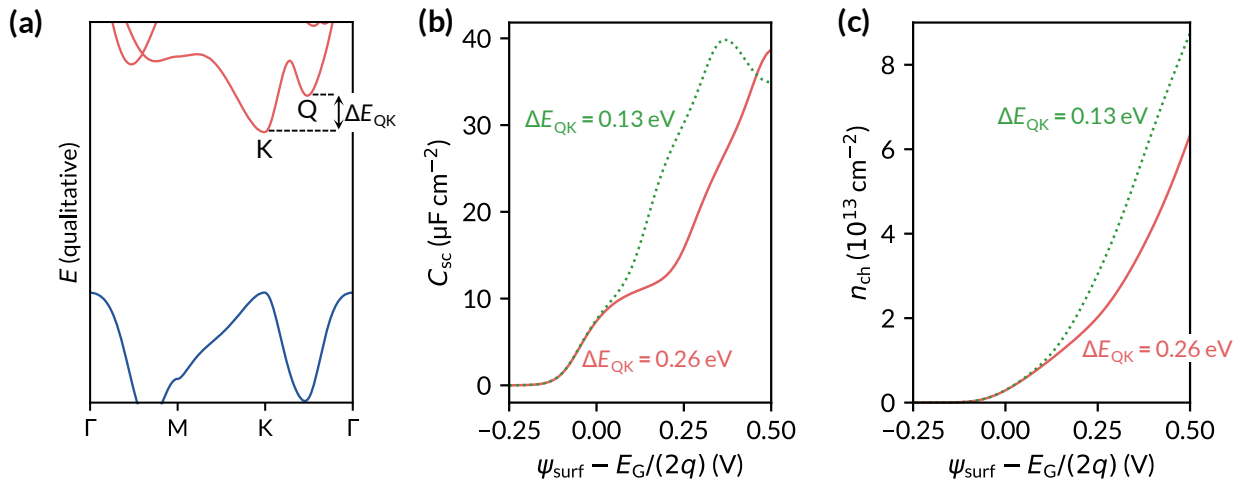

**Figure S1.** (a) Band structure of monolayer MoS₂ near the conduction and valence band edges obtained from DFT, where the energy separation $\Delta E_{QK}$ is labeled between the Q- and K-valleys. Energies ($E$) are not to scale (e.g., the band gap is reduced to highlight details of the conduction band). Note that we neglect spin-orbit coupling in this work, although for monolayer MoS₂, including spin-orbit coupling only negligibly influences the value of $\Delta E_{QK}$ obtained from DFT.[1] (b) Computed semiconductor capacitance $C_{sc}$ and (c) carrier density $n_{ch}$ for $n$-type monolayer MoS₂ capacitors with Q-K energy separations $\Delta E_{QK} = 0.13$ and 0.26 eV.

To obtain a local DOS (LDOS) profile with $\Delta E_{QK} = 0.13$ eV, we take the LDOS used in the main text with $\Delta E_{QK} = 0.26$ eV and splice together the LDOS at energies $E < E_C + 0.13$ eV and $E > E_C + 0.26$ eV to create a continuous LDOS profile with $\Delta E_{QK} = 0.13$ eV. We then compute $C_{sc}$ and $n_{ch}$ with this LDOS using the same methodology as described in the main text.

As shown in Figures S1b and S1c, $C_{sc}$ and $n_{ch}$ are the same for both values of $\Delta E_{QK}$ we consider at low $\psi_{surf}$. However, as $\psi_{surf}$ increases, the states near the Q-valley contribute at lower energies for the LDOS profile with $\Delta E_{QK} = 0.13$ eV, resulting in an earlier onset for the second linear region of $C_{sc}$, thereby increasing $n_{ch}$. Although this lower $\Delta E_{QK}$ shifts this linear region of the $C_{sc}$ curve to the left, it does not significantly affect the maximum value of $C_{sc} \approx 40$ μF/cm² in the range of $\psi_{surf}$ we consider.

Next, we repeat the calculations of $C_G$ and $n_{ch}$ presented in Figures 4a,b of the main text at EOTs of 0.5, 1, and 2.5 nm using $\Delta E_{QK} = 0.13$ and $0.26$ eV. As shown in Figures S2a and S2b, the value of $\Delta E_{QK}$ used affects neither $C_{sc}$ nor $n_{ch}$ at EOT = 2.5 nm since $C_G$ is dominated by the oxide capacitance $C_{ox}$ at sufficiently large EOTs. At EOT = 0.5 and 1 nm, however, we find that the higher $C_{sc}$ at $\Delta E_{QK} = 0.13$ eV allows $C_G$ and $n_{ch}$ to grow closer to the classical limit compared to $\Delta E_{QK} = 0.26$ eV. This result signifies that smaller $\Delta E_{QK}$ provides further advantage of monolayer MoS₂ over Si and III-V channels from the point of view of quantum capacitance and channel carrier density at a given overdrive voltage $V_{OV} = V_G - V_T$. In practice, note that $\Delta E_{QK}$ is controlled by the strain applied[3] and may be controlled by the environmental dielectric as well.

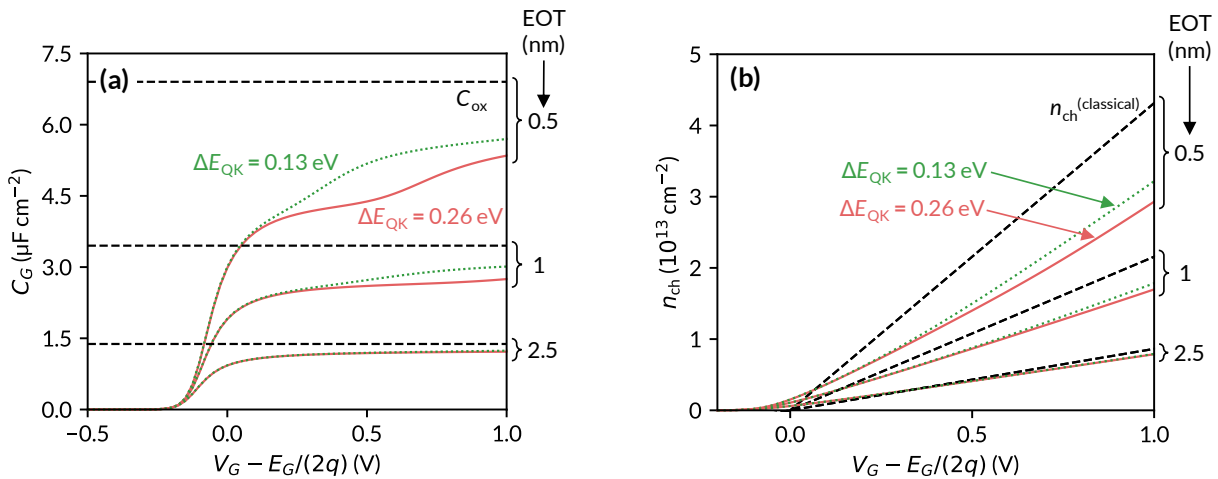

**Figure S2.** (a) Gate capacitance $C_G$ and (b) carrier density $n_{ch}$ for $n$-type MoS₂ with Q-K energy separations $\Delta E_{QK} = 0.13$ and $0.26$ eV at EOTs of 0.5, 1, and 2.5 nm. Dotted green (solid red) lines represent the LDOS profile obtained using $\Delta E_{QK} = 0.13$ (0.26) eV. Dashed lines mark the oxide capacitance $C_{ox} = 3.9\epsilon_0/$EOT and the conventionally calculated carrier density $n_{ch}^{classical} = C_{ox}(V_G - V_T)/q$, where the threshold voltage is $V_T = E_G/(2q)$.

## S2. Comparison to Quantum Capacitance Approximation

As discussed in the main text, the classical approximation of gate capacitance, $C_G = C_{ox}$ (where $C_{ox}$ is the oxide capacitance) and the classical approximation for charge carrier density, $n_{ch}^{classical} = C_{ox}(V_G - V_T)/q$, are well-known to be inaccurate near or below the threshold voltage $V_T$. Instead, $C_G$ is typically modeled in 2D semiconductors as the series combination of the quantum capacitance $C_q$ and the oxide capacitance $C_{ox}$, yielding $C_G^{-1} \approx C_{ox}^{-1} + C_q^{-1}$. Since $C_q$ is a function of the semiconductor's surface potential, when using this equation, $C_G$ must be solved iteratively such that the voltage drop across the oxide and semiconductor are self-consistent with $C_q$. Then, $n_{ch}$ may be approximated at any $V_G$ by integrating this result to obtain the carrier density,

$$n_{ch}^{Cq-corrected} \approx \frac{1}{q} \int_{-E_G/2q}^{V_G} \left[ C_q^{-1}(V_G') + C_{ox}^{-1} \right]^{-1} dV_G'. \qquad (S1)$$

As shown in Figure S3 below, the approximation $C_G^{-1} \approx C_{ox}^{-1} + C_q^{-1}$ matches the rigorously calculated $C_G$ presented in the main text in the subthreshold region (here $|V_G| < E_G/2q$). However, this approximation considerably overestimates the $C_G$ of devices with EOT $\leq 1$ nm at larger overdrive voltages. This finding is consistent with our previous result from Figure 3b in the main text, which shows that $C_q$ similarly overestimates the semiconductor's capacitance in the on-state.

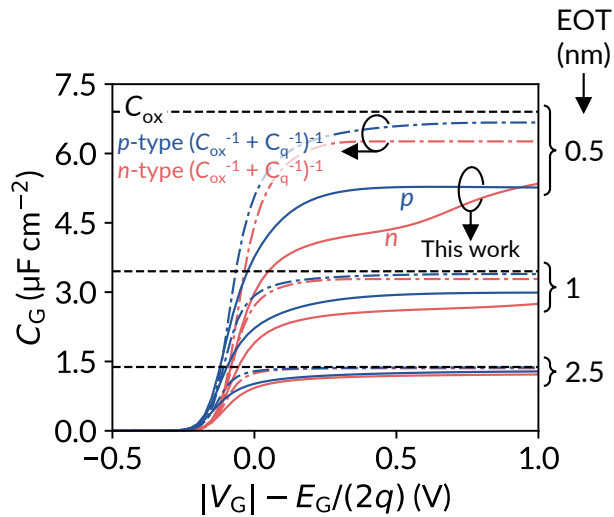

**Figure S3**: Gate capacitance $C_G$ for $n$-type and $p$-type monolayer MoS$_2$ as functions of $V_G$ at EOT = 0.5, 1, and 2.5 nm. Solid red (blue) lines represent the $C_G$ of $n$-type ($p$-type) MoS$_2$ calculated using the full *ab initio* approach described in the main text that includes charge centroid effects. Dash-dotted red (blue) lines are the approximation $C_G^{-1} \approx C_{ox}^{-1} + C_q^{-1}$ (where $C_q$ is calculated using the self-consistent approach described above) which does not include centroid effects, and black dashed lines mark the oxide capacitance, $C_{ox}$. The discrepancies between solid lines and approximations highlight the importance of quantum *and* charge centroid effects in the on-state, which are most important at small EOTs.

Similarly, to compare our rigorously computed $n_{ch}$ in the main text to the carrier density corrected with only the quantum (not centroid) capacitance, we plot our calculated $n_{ch}$ from the main text alongside Equation S1 in Figure S4, below. For a more thorough comparison, we also include our original $n_{ch}^{classical}$ on the same plot. At an EOT of 2.5 nm, Equation S1 accurately approximates our rigorously calculated $n_{ch}$ in both the off- and on-states. However, at EOT = 0.5 and 1 nm, Equation S1 does not correctly predict the charge in the on-state significantly better than the classical approximation $n_{ch}^{classical}$. Again, this result can be understood based on Figure 3b in the main text, where we show that our rigorously calculated semiconductor capacitance $C_{sc} < C_q$ in the on-state. From these results, we conclude that although including corrections for $C_q$ enables good approximations of $C_G$ and $n_{ch}$ in and near the off-state, the more rigorous approach for calculating these quantities (i.e., including spatial variations in the density of states, potential, and volumetric charge density) presented in the main text should be used to understand 2D semiconductor devices with sub-1 nm EOT in the on-state.

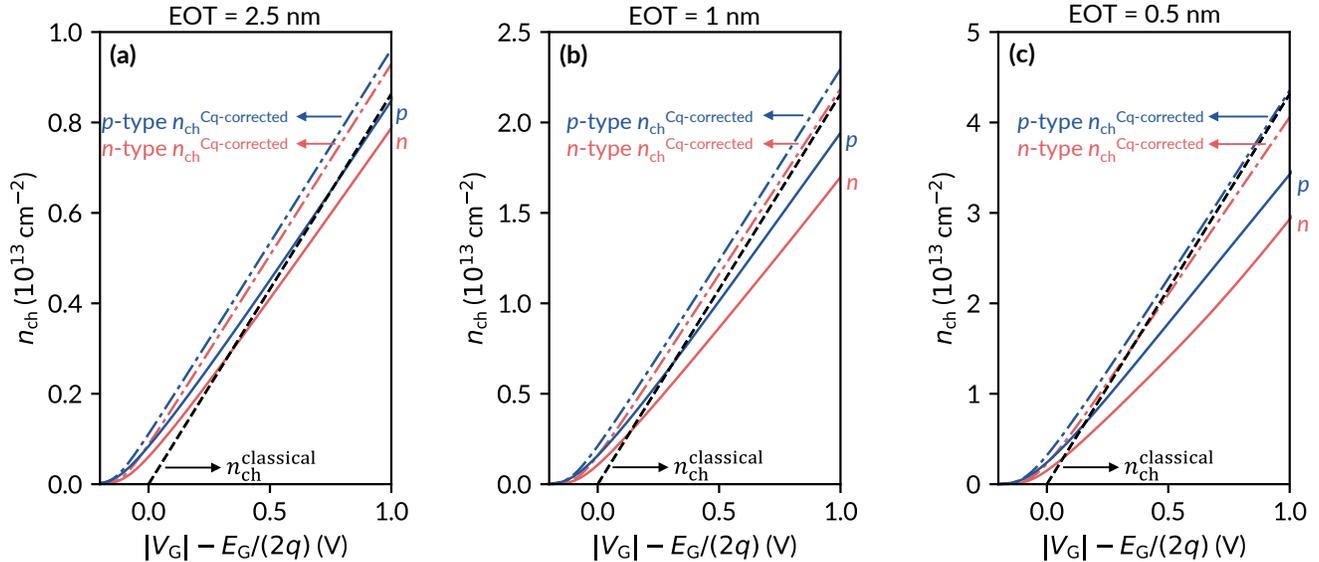

**Figure S4**: Charge carrier density $n_{ch}$ for *n*- and *p*-type monolayer MoS$_2$ as functions of $V_G$ at (a) EOT = 2.5 nm, (b) EOT = 1 nm, (c) EOT = 0.5 nm. Solid red (blue) lines represent $n_{ch}$ of *n*- (*p*-) type MoS$_2$ calculated using the full *ab initio* approach described in the main text, which includes charge centroid effects. Dashed red (blue) lines represent the quantum capacitance-corrected carrier density $n_{ch}^{Cq-corrected}$ (Equation S1, using the self-consistent approach described above) that does not include centroid effects, and black dashed lines mark the classically calculated charge carrier density, $n_{ch}^{classical} = C_{ox}(V_G - V_T)/q$. The discrepancies between solid lines and approximations highlight the importance of quantum *and* charge centroid effects in the on-state, which are most important at small EOTs.

## S3. Monolayer MoS$_2$ Dual-Gated Capacitors

We calculate $n_{ch}$ and $C_G$ of dual-gated monolayer MoS$_2$ capacitors by solving equations (4) and (5) in the main text self-consistently, just as we do when computing these quantities for single-gated devices. Here we use the device schematic shown in Figure S5a, which is similar to the single-gated device in Figure 1b except that the relative permittivities of both the top and bottom insulator are set to 20 and the thicknesses of the top and bottom insulators are identical. We update the bottom boundary condition [previously $\partial V(z)/\partial z = 0$ at this boundary for single-gated devices] to $V(z) = V_G$, where $V_G$ is the gate voltage applied at both the top and bottom electrodes. We plot $C_G$ and $n_{ch}$ at EOT = 0.5, 1, and 2.5 nm for both *n*-type and *p*-type monolayer MoS$_2$ for this dual-gated device in Figures S5b,c.

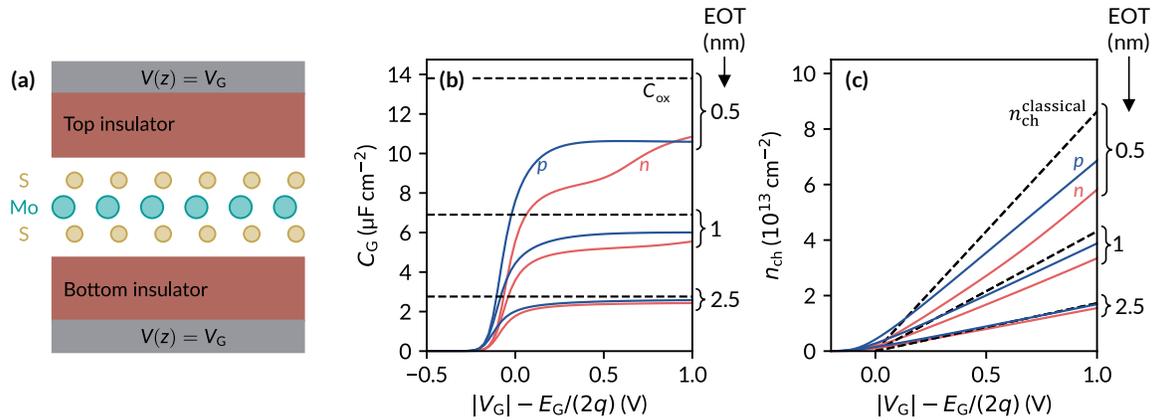

**Figure S5.** (a) Schematic of a monolayer MoS$_2$ dual-gated MOS capacitor with boundary conditions applied when solving equations (4) and (5). (b) Gate capacitance $C_G$ and (c) carrier density $n_{ch}$ for these dual-gated devices. Solid red (blue) lines represent $n$-type ($p$-type) MoS$_2$. Dashed lines mark the total dual-gated oxide capacitance $C_{ox} = 2 * (3.9\epsilon_0/\text{EOT})$ (where the prefactor of 2 accounts for both gates) and the conventionally calculated carrier density $n_{ch}^{classical} = C_{ox}(V_G - V_T)/q$, where the threshold voltage is $V_T = \pm E_G/(2q)$ for $n$- and $p$-type devices, assuming the same gate metal.

## SUPPORTING REFERENCES:

1    Gaddemane, G., Gopalan, S., Van de Put, M. L. & Fischetti, M. V. Limitations of ab initio methods to predict the electronic-transport properties of two-dimensional semiconductors: the computational example of 2H-phase transition metal dichalcogenides. *J Comput Electron* **20**, 49-59 (2021). DOI: 10.1007/s10825-020-01526-1

2    Ryou, J., Kim, Y.-S., Kc, S., & Cho, K. Monolayer MoS$_2$ Bandgap Modulation by Dielectric Environments and Tunable Bandgap Transistors. *Sci Rep* **6**, 29184 (2016). DOI: 10.1038/srep29184

3    Hosseini, M., Elahi, M., Pourfath, M., Esseni, & D. Strain induced mobility modulation in single-layer MoS$_2$. *J Phys D: Appl. Phys* **48**, 375104 (2015). DOI: 10.1088/0022-3727/48/37/375104